\newcommand{\ttau}{\tilde{\tau}}
\documentclass[aps,pre,twocolumn,showpacs]{revtex4}

\usepackage{epsfig}
\usepackage{graphicx}
\usepackage{amsmath}
\usepackage{amssymb}
\usepackage{amsfonts}
\usepackage{dcolumn}

\draft

\begin{document}
\title{Dark solitons as quasiparticles in trapped condensates}
\author{V. A. Brazhnyi$^{1}$}
\email{brazhnyi@cii.fc.ul.pt}

\author{V. V. Konotop$^{1,2}$} 
\email{konotop@cii.fc.ul.pt}

\author{L. P. Pitaevskii$^{3}$}
\email{lev@science.unitn.it}

\address{$^{1}$Centro de F\'{\i}sica Te\'orica e Computacional,
Universidade de Lisboa, Complexo Interdisciplinar, Avenida
Professor Gama Pinto 2, Lisboa 1649-003, Portugal
\\
$^2$ Departamento de F\'{\i}sica,
Universidade de Lisboa, Campo Grande, Edif\'{\i}cio C8, Piso 6, Lisboa
1749-016, Portugal
\\
$^{3}$Dipartimento di Fisica, Universit\`{a} di Trento and
Istituto Nazionale per la Fisica della Materia, CRS-BEC, 38050
Trento, Italy, and Kapitza Institute for Physical Problems, 119334
Moscow, Russia }

\begin{abstract}
\par
We present a theory of dark soliton dynamics in trapped quasi-one-dimensional Bose-Einstein condensates, which is based on the local density approximation.
The approach is applicable for arbitrary polynomial nonlinearities of the mean-field equation governing the system as well as to arbitrary polynomial traps.
In particular, we derive a general formula for the frequency of the soliton oscillations in confining potentials. 
A special attention is dedicated to the study of the soliton dynamics in adiabatically varying traps.
It is shown that the dependence of the amplitude of oscillations {\it vs} the trap frequency (strength) is given by the scaling law
$X_0\propto\omega^{-\gamma}$ where the exponent $\gamma$ depends  on the type of the two-body interactions, on the exponent of the polynomial confining potential, on the density of the condensate and on the initial soliton velocity.
Analytical results obtained within the framework of the local density approximation are compared with the direct numerical simulations of the dynamics, showing remarkable match. 
Various limiting cases are addressed. 
In particular for the slow solitons we computed a general formula for the effective mass and for the frequency of oscillations.

\end{abstract}
 \maketitle


\section{Introduction}

One of the main properties of solitons, making them
to be of special interest for physical applications,
is preserving their localized shapes during evolution
and mutual interactions~\cite{soliton}.
Due to this robustness solitons can be regarded as
quasiparticles and systems possessing
large number of such excitations can be described
in terms of the distribution function governed
by the kinetic equation~\cite{kinetic}.

In the mean-field theory~\cite{Pit1}
description of the quasi-one-dimension homogeneous
Bose gas is reduced to the  exactly
integrable nonlinear Shr\"{o}dinger (NLS)  
[or one-dimensional (1D) Gross-Pitaevskii (GP)] equation,
and therefore  solitons are expected to play
a prominent role in the dynamical and statistical
properties of low-dimensional condensates.
When interatomic interactions are repulsive,
the GP equation possesses dark (or grey) soliton
solutions~\cite{Tsuzuki,Pit1,termin}. 
Existence of the dark solitons was confirmed by a
number of recent experiments with BEC's confined by elongated traps~\cite{experim}.

In practice, condensates appear to
be never homogeneous, and therefore effect of
external potentials on the dark-soliton dynamics
is a subject of special interest
(see e.g.~\cite{FMS,Anglin,BK1,KP}
and references therein). An inhomogeneity
of a system by itself does not invalid possibilities
of description of solitons as quasiparticles
(in some approximation, of course).
In particular, one can explore Hamiltonian approach to  an effective
particle with one degree of freedom, instead of dealing with the original
equation for the macroscopic wave function,
which is a system with infinite degrees of freedom.
Moreover, one can extend the respective description
on the gas of solitons, which now will be
described by a distribution function governed either
by Fokker-Planck equation (for the case
where a soliton bearing system interacts
with a thermal bath, see e.g.~\cite{KV})
or by a kinetic equation with respective
collision integral, as this is shown in~\cite{FMS} for the case of interaction of
solitons with a noncondensed atoms.

A quasiparticle description of dark solitons can be
obtained from the perturbation theory
in adiabatic approximation~\cite{BK1}
(sometimes called the collective variable approach).
At the same time, as was shown in~\cite{KP},
a concept of a quasiparticle naturally emerges from the
Landau theory of superfluidity and can be justified
on the basis of the mean field theory
within the framework of the {\em local density
approximation}.  It turns out, that a dark
soliton moves in an external potential without
deformation of its density profile as a particle
of mass 2$m$. The local density approximation  is rather general, 
allowing direct extension to other nonlinear equations, related to the BEC dynamics,
as well as to various (non-parabolic) types of the
trap potential.  Building up such a generalized theory
is the main goal of the present paper.

In real experimental conditions the external trap potential can depend not only on coordinate, 
but also on time. That is why the second aim of the present work is the description of the effect of adiabatic
time-dependence of the external parameters  
on the dark soliton motion.

The paper is organized as follows. We start with the dynamics of a dark soliton
in an adiabatically changing parabolic trap (Sec.~\ref{sec:DarkSoliton}).
In Sec.~\ref{sec:GeneralApproach}, we develop our  Hamiltonian theory
for solitons described by generalized polynomial NLS
equations and show  how such approach is related
to the mean field approximation.
In Sec.~\ref{sec:Examples} we  consider in detail
examples of dark soliton dynamics, which include
the cases of non-parabolic trap and models
with higher nonlinearity.  The consideration is provided
within the framework of the local density
approximation and is verified by direct numerical
simulations of the dark soliton dynamics.
In this section we also show how one can
modify the perturbation theory for dark
solitons to take into account adiabatic
change of the trap frequency (Sec.~\ref{sec:perturbed})
and make comments on the dynamics of small
amplitude dark solitons (Sec.~\ref{sec:smallamp}).
The outcomes are summarized in the Conclusion and
technical details of some calculations are given in
the Appendices.

\section{Dark soliton in  a time-depended
parabolic trap}
\label{sec:DarkSoliton}

Let us start with the dynamics of a dark
soliton described by the GP equation
\begin{eqnarray}
\label{NLS_GP}
i\hbar\Psi_t=-\frac{\hbar^2}{2m}\Psi_{xx}+
\frac{1}{2} m\omega^2x^2\Psi+g|\Psi|^{2}\Psi-\mu\Psi.
\end{eqnarray}
Here $g=2\hbar ^2a_s/(ma_{\perp}^2)$,
$a_s$ is s-wave scattering length and
$a_\perp$ the transverse linear oscillator length,
which describes the BEC in an elongated trap at low densities~\cite{Pit1} (see also~\cite{BK1} for the details of derivation by means of the multiple scale expansion method).

It has been shown in Ref.~\cite{KP} (see also the details
below Sec.~\ref{sec:GeneralApproach}) that the dark
soliton dynamics in a parabolic trap can be successfully
described within the framework of
the local density approximation.
This means that, in spite of the presence of the
trap, one starts with the solution of
the 1D homogeneous (i.e. when $\omega=0$)
GP equation~\cite{Tsuzuki}
(see also~\cite{Pit1},  \S 5.5):
\begin{eqnarray}
\label{GPsoliton}
\Psi \left( x,t \right ) =\sqrt{n_0}
\left( i\frac{v}{c}+ \frac{\sqrt{c^2-v^2}}{c}
\tanh \left[ \frac{x-X(t)}{\ell} \right] \right)
  \;,
\end{eqnarray}
where $X(t)=vt$, $v$ is the velocity of the soliton,
$n_0$ is the unperturbed linear density,
$c=\sqrt{gn_0/m}$ is the speed of sound
and $\ell=\hbar/(m\sqrt{c^2-v^2})$ is the width of
the soliton. Then the influence of the trap is
accounted by considering a general
function $X(t)$ which dependence on time is to be obtained.

The energy of the system can be defined as
\begin{eqnarray}
\label{energyGP}
 E=\int\left[\frac{\hbar ^{2}}{2m}\left|
 \Psi _{x}\right| ^{2}+\frac{g}{2}
\left( |\Psi |^2-n_0\right)^2 \right]dx \nonumber \\
=\frac{4}{3}\hbar c n_0
\left ( 1-\frac{v^2}{c^2}\right )^{3/2}
\end{eqnarray}
and for the dark soliton solution (\ref{GPsoliton})
can be rewritten in a form of the conservation law
 \begin{eqnarray}
 \label{disp_rel}
 c^2\left( X\right)-v^{2}=({\cal G}E)^{2/3}
\end{eqnarray}
where ${\cal G}=3g/(4\hbar m)$.
The introduced dependence $c=c(X)$ is the key
point of the local density approximation:
the sound velocity is substituted by its
local value computed in the point where
the center of the soliton is located.
In the Thomas-Fermi (TF) approximation,
when the atomic density is given
by $n(x)=\frac 1g \left(\mu-\frac{m\omega^2x^2}{2}\right)$,
one has
\begin{eqnarray}
\label{cGP}
c^2(X)=\frac gm n(X)=c_0^2-\frac 12 \omega^2X^2
\end{eqnarray}
with $c_0=\sqrt{\mu/m}$. Substituting  $v=dX/dt$ in (\ref{disp_rel}),
the energy conservation can be rewritten as follows~\cite{KP}
\begin{eqnarray}
\frac{m_s}{2}\left( \frac{dX}{dt}\right) ^{2}+
\frac {m_s\omega_s^2}{2}X^2
=E_*\;.
\label{final}
\end{eqnarray}
Here we introduced the effective mass of
the soliton considered as a  quasiparticle
\begin{eqnarray}
\label{eff_mass_GP}
m_s=2m\,,
\end{eqnarray}
the frequency of the soliton
oscillations $\omega_s=\omega/\sqrt{2}$~\cite{Anglin,BK1,KP,PFK,com}
and the effective soliton energy
 \begin{eqnarray}
 \label{E0}
  E_*=\frac{m_sc_*^2}{2},\qquad c_*^2=c_0^2-({\cal G} E)^{2/3}
 \end{eqnarray}
which, altogether with $E$, is a constant of motion.
The amplitude of oscillations governed by (\ref{final}) is
\begin{eqnarray}
\label{X0}
X_0=\sqrt{2E_*/m_s\omega_s^2}.
\end{eqnarray}

One of the characteristic features of the
introduced quasiparticles is that their
dynamics is determined not only by their
local properties (velocity and amplitude)
but also by the environment, i.e.
by the unperturbed density.
As a result any change of the trap
characteristics (say, trap frequency or geometry)
will affect solitons not only by changing the
domain of their motion  but also through
the change of the density. It turns out that the local density approximation is a
 suitable framework for description of  mentioned phenomena in the case when time
  variation of the parameters of the system is slow enough.
  
According to a general law of the Hamiltonian mechanics, the adiabatic invariant
\begin{eqnarray}
\label{AIint}
I\left(E\right) =\frac{1}{2\pi }\oint pd X  \label{I}
\end{eqnarray}
stays constant~\cite{LL_mech}. Time dependence
of the amplitude of oscillation can be defined from this condition.
The canonical momentum, which enters in (\ref{AIint}),
 can be computed
explicitly using the formula
\begin{eqnarray}
\label{p_can}
        p=\int_0^v\frac{\partial E}{\partial v}\frac{dv}{v}\,
\end{eqnarray}
what gives
\begin{eqnarray}
p=-2n\hbar\left( \frac{v}{c}
\sqrt{1-\frac{v^{2}}{c^{2}}}+\arcsin \left( \frac{v}{c}
\right) \right)\;.  \label{p}
\end{eqnarray}
It turns out, however, for calculation of
the adiabatic invariant it is more convenient to use the
general equation between $I$ and the frequency of oscillations:
\begin{eqnarray}
\label{period}
\frac{dI}{dE}=
\frac{1}{2\pi }\oint \frac{dX}{v}=(\omega_s)^{-1}
\;.
\end{eqnarray}

Taking into account that $\omega_s=\omega/\sqrt{2}$
 does not depend on $E$ and using an
 obvious boundary condition $I=0$ at $v\to 0$, we easily
 find a simple equation
 \begin{equation}
I=\frac{\sqrt{2}}{\omega }\left( E-\frac{4\hbar m}{3g}%
c_{0}^{3}\right) \;.
\label{I_dark}
\end{equation}
It is not difficult to show
(see, for example, Eq. (17.10) in \cite{Pit1}),
 that in the TF approximation one has
 \begin{equation}
 c_{0}^{2}=gn(0)/m
\propto \omega^{2/3},\quad\mbox{i.e.}\quad c_{0}^{3}\propto \omega\;, 
\label{c0}
\end{equation}
so the second term on r.h.s. of (\ref{I_dark}) is constant.

 Thus preserving the adiabatic integral in an
adiabatic process implies preserving $E/\omega$,
what in the case of slowly varying frequency
implies $E\propto\omega$. Taking again into account
that according (\ref{c0}) in the TF approximation $c_0\propto\omega^{1/3}$,
one deduces from (\ref{E0}) that $E_*\propto\omega^{2/3}$.
Finally, the scaling law for the amplitude of oscillations,
defined by (\ref{X0}), reads 
\begin{eqnarray}
\label{x_v_om_harm1}
X_0\propto\omega^{-2/3}.
\end{eqnarray}

It is worth to underline that this law is  different
that one for a conventional harmonic oscillator, where 
$ X_0 \propto \omega^{-1/2},
$ 
even though the motion of the soliton is pure harmonic. The point is that in our case the ratio
$E/\omega$, but not $E_*/\omega $, is preserved.

An important feature of the soliton dynamics is that in the case at hand the soliton
frequency does not depend on the energy. Hence the frequency of the soliton oscillations
does not depend on the amplitude of the soliton,
what corroborates with the analysis of the
oscillations of the small-amplitude solitons [see (\ref{om_small_amp}) below and
subsequent discussion] as well with the earlier studies~\cite{KP,PFK}.

\begin{widetext}

\begin{figure}[h]
\epsfig{file=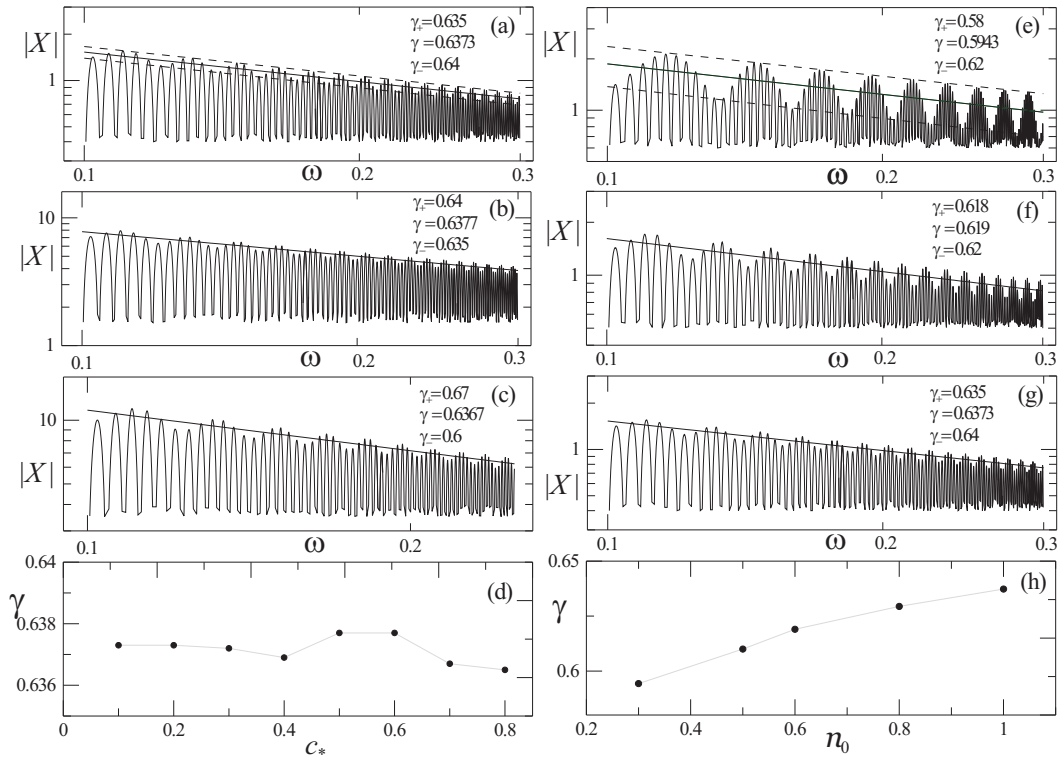,width=14cm}
\caption{Time dependence of the soliton coordinate on the adiabatically changing frequency, modeled by the function  $\omega(t)=(1+0.001t)\cdot 0.1$, in logarithmic coordinates: panels (a)-(c) and (e)-(g). 
Straight solid lines visualize the law $X_0=\tilde{X}_0\omega^{-\gamma}$. 
Dashed lines in panels (a) and (e) show the exponents $\gamma_\pm$ [see (\ref{pm})] while in the rest of panels only the average $\gamma$ is shown. 
In panels (a)-(c) the parameters are $n_0=1$ and $v=0.1; 0.5$ and $0.8$, correspondingly. 
The respective matching parameter is $\tilde{X}_0=0.385; 1.93; 2.6$. 
In panels (e)-(g) the parameters are $v=0.1$, $n_0=0.3; 0.6$, and $1$, correspondingly. 
The respective matching parameter is $\tilde{X}_0=0.622; 0.445; 0.385$.  
In panel (d) we plot dependence of $\gamma$ on the initial soliton velocity $v$ for the case $n_0=1$. 
In panel (h) we depict the dependence of $\gamma$ on the initial density $n_0$ corresponding to the case of the initial velocity $v=0.1$. 
In the numerical calculations we take $\hbar=1$, $m=1$ and $g=1$.}
\label{fig1}
\end{figure}

\end{widetext}

We have checked the obtained predictions,
made on the basis of the local density approximation, numerically.
The typical results are presented in Fig.~\ref{fig1}.

The local density approximation essentially uses that the
background of the condensate is static, i.e. that the dark soliton
motion does not excite the motion of the whole condensate. In
practice, due to finiteness of the system, such a supposition
strictly speaking does not hold, and the whole condensate also
undergoes oscillations with the frequency of the condensate
$\omega$, what follows directly from the Ehrenfest theorem. The
difference of the frequencies of the condensate and of the dark
soliton, i.e. between $\omega$ and $\omega_s$, results in the
beating of the dark soliton~\cite{BK1}, which are clearly
observable in Fig.~\ref{fig1}. Respectively, one can identify
the two slopes corresponding to the maxima and to the minima of
the soliton amplitudes. We will use the subindexes ``$+$'' and ``$-$''
for the respective quantities. In other words, each of the
results presented in panels (a) -- (c) and (e) -- (g) is
characterized by the two scaling laws: $X_{0,\pm}
=\tilde{X}_\pm\omega^{-\gamma_\pm}$ shown explicitly in Fig.~\ref{fig1} (a),(e). 
The exponents $\gamma_\pm$
 are different (although their difference is relatively
small), what requires a definition of some averaged exponent
$\gamma$ which could be compared with the theoretical
predictions. We obtain such exponent numerically from the
dynamics of the averaged amplitude, i.e. using the formula
\begin{eqnarray} 
\label{pm}
\frac{\tilde{X}_0}{\omega^{\gamma}}=\frac
12\left(\frac{\tilde{X}_+}{\omega^{\gamma_+}}
+\frac{\tilde{X}_-}{\omega^{\gamma_-}}\right)\,.
\end{eqnarray}
Summary of the results for the averaged exponent
$\gamma$ are presented in panels (d) and (h). As one can see
from the figures the law of the change of
the amplitude of soliton oscillations stays close to the
predicted law $\gamma=2/3$ for relatively slow solitons and
relatively large densities. Meantime deviations are clearly seen
in Fig.~\ref{fig1} (c) and (e). In the last case the exponent
$\gamma$ is essentially less than the predicted in our
analytical consideration. It turns out, however that the
mentioned deviation from $2/3$ law is observed for small
densities. This is natural from point of view  of the theory. Indeed, our
consideration was based on the TF approximation for the atomic
density, when $n_0\propto\omega^{2/3}$. This approximation fails
at low densities, and must be substituted by the Gaussian
distribution, where $n_0\propto\omega^{1/2}$. Then by repeating
the above our arguments for the Gaussian distribution, instead
if the TF one, one finds 
\begin{eqnarray}
\label{x_v_om_harm2}
X_0\propto\omega^{-1/2},
\end{eqnarray}
 i.e. the law
of the dependence of the amplitude of oscillation of the
conventional linear oscillator on the frequency, what
corroborates with the numerical findings.

\section{General approach}
\label{sec:GeneralApproach}

\subsection{Generalized equation.}
The theory developed in the previous section can be generalized for NLS equation
with arbitrary power-law nonlinearity and non-parabolic potential. More specifically,
in the present section we consider the equation
\begin{eqnarray}
\label{GP}
i\hbar\Psi_t=-\frac{\hbar^2}{2m}\Psi_{xx}+
U(x)\Psi+g|\Psi|^{2\alpha}\Psi-\mu\Psi
\end{eqnarray}
where $\alpha$ is a positive integer and $g>0$,
which describes interacting particles of mass $m$ in an
external potential $U(x)$. The exponent $\alpha$
characterizes the effective inter-particle interactions.
In particular when
$\alpha=1$ and $g=2\hbar
^2a_s/(ma_{\perp}^2)$ one recovers the GP equation
(\ref{NLS_GP})
considered in the previous section.

The chemical potential $\mu$
introduced in (\ref{GP}) is determined by the link
valid for a homogeneous condensate: $\mu=gn^{\alpha}$. Thus the
sound speed $c$ connected to the chemical potential
by the relation $mc^2=nd\mu/dn$ can be expressed as follows
\begin{eqnarray}
\label{speed}
        c^2=\frac{\alpha g}{m}n^{\alpha}\;.
\end{eqnarray}

There are several reasons to consider more general
equation (\ref{GP}). First of all, equation (\ref{NLS_GP}),
being completely integrable, possesses very specific
soliton properties. It is interesting to investigate the soliton dynamics
in a more general situation. 
The case of $\alpha=2$ is particularly important, because
corresponding equation can be used in different physical
problem. Such a situation can take place near the Feshbach
resonance. In this case the s-wave scattering length depends on
magnetic field as $a_s=a_g+\Delta/(B-B_0)$ where $a_g$ is the background value of the scattering length,
and $B_0$ and $\Delta$ are the location and width of the resonance.
If magnetic field is equal to
$B_c=B_0-\Delta/a_g$, the  scattering length turns to zero and the dominant
interaction among atoms is due to three-body effects.

Indeed, in the higher approximations of the
Bogoliubov theory expansion of the chemical potential
of an uniform gas with respect to density $n$ has form
\begin{eqnarray}
\mu=a_sn \left [b_1+b_2(na_s^3)^{1/2}+
b_3(na_s^3)\ln\frac{1}{na_s^3}\right ]
+g_2n^2\,,
\end{eqnarray}
where $b_1=4\pi\hbar^2/m$ and other coefficients $b$
can be calculated (see \cite{Pit1},  \S 4.2).
Coefficient $g_2$ depends on three-body interactions
and cannot be calculated explicitly. However, it
stays finite for $B=B_c$, while
three first terms disappear, giving $\mu=g_2n^2$~\cite{footnote}.
Correspondingly, the non-linear term in the mean-field equation
has form $g_2\mid \Psi \mid ^4\Psi$. The sign of $g_2$ cannot
be defined from general considerations. We assume that
$g_2>0$.
 After averaging with respect
to the transverse motion we obtain (\ref{GP}) with $\alpha=2$ and
$g=g_2/(3\pi^2a_{\perp}^4)$.

 Another physical system where the equation of the state with $\alpha=2$ is valid, 
is an 1D Bose gas in so-called Tonks-Girardeau  (TG) limit of inpenetrable 
particles. This limit can be achieved for a gas of small density. It has 
been shown by Girardeau that there exists an exact mapping between states
of  this system and an {\it ideal} 1D Fermi gas. In particular in this 
case one has $\mu=gn^2$ with
$g=\hbar^2\pi^2/(2m)$. It has also been rigorously
shown that one can find density distribution
of a such gas in a 1D
trap by minimization of the energy functional~\cite{Lieb}
\begin{eqnarray}
\label{energyTG}
 E=\int\left[\frac{\hbar ^{2}}{2m}\left
 [(\sqrt{n})_x \right]^{2}+\frac{g}{3}n^3
 +U(x)n \right]dx\,. 
\end{eqnarray}

On the basis of these considerations authors of Ref. [18] suggested to 
use equation (19) for dynamics of the TG gas. However, the 
hydrodynamic-like equation (19) can not give a satisfactory description of 
dynamics of an ideal Fermi gas. Nevertheless it can be useful for a Bose 
gas near the TG limit, where equation of state approximately follows the 
$\alpha=2$ law, but dynamic  is still not an ideal gas-type.

The case $\alpha=2$ is often referred also as a quintic nonlinear Schr\"{o}dinger (QNLS) equation. For the sake of brevity in what follows use this terminology. 

We mention that other polynomial models are also considered in literature~\cite{Sal}.

\subsection{ Soliton in the generalized equation.}

Let us consider now a condensate in the absence
of external field, $U(x)= 0$. Eq. (\ref{GP})
takes the form
\begin{eqnarray}
\label{NLS}
i\hbar\Psi_t=-\frac{\hbar^2}{2m}\Psi_{xx}+g|\Psi|^{2\alpha}\Psi-\mu\Psi
\end{eqnarray}
and is subject to the finite density boundary conditions:
\begin{eqnarray}
\label{bound}
        \lim_{x\to\pm\infty}\Psi(x,t)=\sqrt{n_0}e^{\pm i\theta}
\end{eqnarray}
where the constant $\theta$ can be considered
without restriction of generality in the
interval $[0,\pi/2]$: $\theta\in [0,\pi/2]$.
Then dark solitons, $\Psi_s(x,t)$, will be
associated with traveling wave solutions,
 characterized by the following dependence of the density
 on the spatial coordinate and time:
\begin{eqnarray}
\label{running}
|\Psi_s(x,t)|^2\equiv \eta^2(x-vt)\,,
\end{eqnarray}
where $v$ is the soliton velocity. Below such solutions will also be referred to as unperturbed.

The energy of the soliton solution can be defined as
\begin{eqnarray}
\label{energy}
 E=\int{\cal E}(x)dx
\end{eqnarray}
where the energy density ${\cal E}(x)$ is given by
\begin{eqnarray}
\label{density}
{\cal E}(x)= \frac{\hbar ^{2}}{2m}\left| \Psi _{x}\right| ^{2}+\frac{g}{
\alpha +1}
\left( |\Psi |^{2\alpha +2}-n_0^{\alpha +1}\right)
\nonumber
\\
-
gn_0^{\alpha }\left( \left| \Psi \right| ^{2}-n_0\right)\,.
\end{eqnarray}
The energy is an integral of motion. Hence taking
into account that the dark soliton depends on
the two parameters $(n_0,v)$ and connecting the mean  density with the speed of sound by (\ref{speed}), one concludes that the energy of the dark soliton is a function of $c$ and $v$:
\begin{eqnarray}
\label{cons_en}
E=E(c,v)\,.
\end{eqnarray}

\subsection{The local density approximation.}

Consider now propagation of a dark soliton in a condensate with the density, varying due to the external trap potential: $n=n(x)$ with $n(0)=n_0$ (for the sake of definiteness the trap potential will be assumed having minimum at $x=0$: $U(0)=0$). In particular, in the TF approximation the function $n(x)$ is given by
\begin{eqnarray}
\label{TF}
        n(x)= n_{TF}(x)\equiv g^{-1/\alpha}[\mu-U(x)]^{1/\alpha}\,.
\end{eqnarray}
This formula determines the dependence of the sound velocity on the spatial coordinate [c.f. (\ref{cGP})]:
\begin{eqnarray}
        \label{cx}
        c^2(x)=c_0^2-\frac{\alpha}{m}U(x)\,,
\end{eqnarray}
where $c_0$ is expressed through $n_0$ by the link (\ref{speed}).

Now we define the {\em local density approximation} as an assumption  that the  conservation law (\ref{cons_en}) is valid for a soliton in the inhomogeneous
condensate, i. e.,  that $c$ can be changed to its local value
$c\left( X\right)$, where $X$ is the position of the center of the
soliton, computed using the unperturbed soliton wave function $\Psi_s(x,t)$. Respectively, $X$ and $v$ are considered as functions of time related by the equation
$dX/dt=v(t)$.

Thus, in the local density approximation the equation of motion of the soliton is  determined from (\ref{cons_en}):
\begin{eqnarray}
\label{din_eq_1}
        E(c(X),v)-E=0.
\end{eqnarray}
Here $E$ is the constant energy of the soliton.

Eq. (\ref{din_eq_1}) can be viewed as an equation of motion of a quasiparticle, which can be associated to the dark soliton. Then $E(c(X),v)$ must be associated with the Hamiltonian of the quasiparticle after expressing the velocity $v$ through the canonical momentum $p$ according to the formula (\ref{p_can}).
After inverting this formula, one obtains the Hamiltonian of the quasiparticle:
\begin{eqnarray}
        H(p,X)\equiv E(c(X), v(p,X)).
\end{eqnarray}

Finally, the adiabatic invariant and the frequency are computed according to formulas (\ref{AIint}) and (\ref{period}), which are obviously valid in the general case.

\subsection{Justification of the local density approximation.}

In the present subsection we show that equation for the
energy of a soliton, obtained for an uniform condensate, is
actually valid also for a trapped condensate in the local density
approximation. Thus the trapping potential does not enter
explicitly in the expression for the energy of a soliton.

To this end we define a real-valued wavefunction of the background $F(x)$ such that $n_0(x)=n_0F^2(x)$ is the density of the condensate in the absence of the soliton and  $F(x)$ solves the equation~\cite{BK1}
\begin{eqnarray}
\label{eq:F}
        -\frac{\hbar^2}{2m}F_{xx}+gn_0^\alpha F^{2\alpha+1}+[U(x)-\mu]F=0
\end{eqnarray}
where $n_0=n_0(0)$ subject to the normalization conditions $F(0)=1$ and $F_x(0)=0$.

The density of the "grand canonical energy"  of the inhomogeneous condensate can be written down as follows
\begin{eqnarray}
{\cal E}^\prime (x)=\frac{\hbar^2}{2m}|\Psi_x|^2
+\frac{g}{1+\alpha}n^{\alpha+1}(x)+[U(x)-\mu]n(x)\,.
\nonumber \\
\label{ex}
\end{eqnarray}
Here $n(x)=|\Psi|^2$. Let the  soliton center be at $x=X$, $\ell$ be a soliton width and $L_0$ be the spatial extension of the condensate. Then we introduce $\delta$ such that $L_0\gg \delta \gg \ell$ and separate the integration
on two domains
\begin{eqnarray}
E^\prime=\int\limits_{|x-X|>\delta} {\cal E}^\prime dx +\int\limits_{|x-X|<\delta}
{\cal E}^\prime dx\,.
\label{E12}
\end{eqnarray}
Next, we add to the first term an integral
$\int_{|x-X|<\delta} {\cal E}_{0}^\prime dx $ and, correspondingly,
deduct it from the second term in $E^{\prime}$.

For the case of a dark soliton solution, which is exponentially localized around $x=X$, the first integral can be approximated (with the exponential accuracy) as follows
\begin{eqnarray}
        \int\limits_{|x-X|>\delta} {\cal E}^\prime dx+\int\limits_{|x-X|<\delta} {\cal E}_{0}^\prime dx\approx \int_{-\infty}^{\infty}{\cal E}_{0}^\prime dx=E_0
\end{eqnarray}
where $E_0$ is the energy of the unperturbed condensate.

In order to compute the other two integrals we represent
\begin{widetext}
\begin{eqnarray}
\label{energy_full}
        {\cal E}^\prime(x)-{\cal E}_0^\prime(x)=\frac{\hbar^2}{2m}|\Psi_x|^2+\frac{g}{1+\alpha}\left[n^{\alpha+1}(x)-n_0^{\alpha+1}(x)\right]-gn_0^\alpha (x)\left[n(x)-n_0(x)\right]+\frac{\hbar^2}{2m}\frac{F_{xx}}{F}n(x)-\frac{\hbar^2n_0}{4m}\left(F^2\right)_{xx}\,.
\end{eqnarray}

As it is shown in Appendix~\ref{app:F} the last two terms can be made as small as necessary by choosing the potential large enough, while in the rest of the terms related to the background $x$ can be securely substituted by $X$ (due to their smoothness in the region of the soliton motion). This leads us to the final expression for the energy of the soliton:
\begin{eqnarray}
\label{energy_inhom}
E_s=\int\limits_{|x-X|<\delta} \left({\cal E}^\prime- {\cal E}_{0}^\prime \right) dx\approx
\int\limits_{|x-X|<\delta}\left\{ \frac{\hbar ^{2}}{2m}\left| \Psi _{x}\right| ^{2}+\frac{g}{\alpha +1}
\left[ |\Psi |^{2\alpha +2}-n_0^{\alpha +1}(X)\right]
-gn_0^{\alpha}(X)\left[ \left| \Psi \right| ^{2}-n_0(X)\right] \right\}dx \,.
\end{eqnarray}
\end{widetext}

The obtained integral does not depend (in the leading order) on the particular choice of the parameter $\delta$. Then comparing the expression (\ref{energy_inhom}) with (\ref{energy}), (\ref{density}) one can verify that they lead to the same expression for the soliton energy, where the only substitution $n_0$ by $n_0(X)$ must be made.

\section{Examples of Landau dynamics of dark solitons}
\label{sec:Examples}

In the present section we consider two examples relevant in different ways to the BEC dynamics in low dimensions.

\subsection{Dark soliton of the GP equation
in a polynomial trap.}

\subsubsection{General approach.}

Let us now turn to the case where the ``polynomial'' trap
\begin{eqnarray}
        \label{plynom}
        U(x)=\frac m2 \omega^{2r}x^{2r}
\end{eqnarray}
with $r$ being a positive integer, $r=1,2,...$, and $\omega$ being a function slowly depending on time: $\omega=\omega(t)$. If $r=1$, then $U(x)$ is transformed in the conventional parabolic trap considered in Sec.~\ref{sec:DarkSoliton}. Then $\omega$ is the trap frequency. For this reason and for the sake of brevity of notations in what follows $\omega$ is referred to as a frequency independently on the value of $r$.

The question we are interested in is the dependence of the amplitude of the soliton oscillations on the frequency, subject to the adiabatic change of the last one.
The explicit form of $p$, given by (\ref{p}), allows one to solve the problem analytically in a general case, i.e. for the arbitrary integer $r$.

Now the link between the velocity and the coordinate (\ref{disp_rel}) reads
\begin{eqnarray}
\label{link_vx}
        v^2+\frac 12 \omega^{2r}X^{2r}=c_*^2
\end{eqnarray}
[$c_*$ was defined in (\ref{E0})] and expression for the amplitude of the oscillations of the soliton, $X_0$ is given by:
\begin{eqnarray}
        \label{x0}
        X_0=\frac{2^{1/(2r)}c_*^{1/r}}{\omega}\,.
\end{eqnarray}

Next one can compute the following quantities:

 -- The normalization condition
\begin{eqnarray}
\label{N}
        N=\int_{-x_{TF}}^{x_{TF}}n(x)dx=\frac{2r(2n_0)^{1+1/(2r)}}{2r+1} \left(\frac gm \right)^{1/(2r)} \frac 1\omega\,,
\end{eqnarray}
where $N$ is the total number of atoms and we introduced the TF radius
\begin{eqnarray}
x_{TF}=\left(\frac{2gn_0}{m}\right)^{1/(2r)}\frac{1}{\omega}\,.
\end{eqnarray}

-- The adiabatic invariant
\begin{eqnarray}
        \label{AI_GP}
        I= \frac{\hbar m^{1-1/(2r)}c_*^{2+1/r}}{g \omega}G_r\,,
\end{eqnarray}
where the constant $G_r$ is defined in (\ref{Gr}) and the details of calculations are presented in Appendix~\ref{ap:adint}.

-- The frequency of the soliton [using (\ref{period})]
\begin{eqnarray}
\label{frequencyGP}
        \omega_s=R_rc_*^{1-1/r}\omega\,,
\end{eqnarray}
where
\begin{eqnarray}
\label{Rr}
        R_r=\frac{\pi}{2^{1/(2r)}}\left(\int_{-1}^1\frac{dx}{\sqrt{1-x^{2r}}}\right)^{-1} .
\end{eqnarray}

The obtained relations, as well as constancy of the total number of particles $N$ and of the adiabatic invariant $I$ subject to slow change of the frequency readily allow one to get the scaling relations [they follow from (\ref{N}) and (\ref{AI_GP}), respectively]:
\begin{eqnarray}
        \label{scaling}
        n_0\propto\omega^{2r/(1+2r)}\quad\mbox{and}\quad
        c_*\propto \omega^{r/(1+2r)}\,.
\end{eqnarray}
Finally, taking into account the link (\ref{x0}) we arrive at the general scaling relation determining the dependence of the amplitude of the soliton oscillations on the frequency
\begin{eqnarray}
        \label{x_vs_o}
        X_0\propto\omega^{-\gamma},\qquad \gamma=\frac{2r}{1+2r}
\end{eqnarray}

\subsubsection{GP dark soliton in an $x^4$-trap.}

Let us consider in more details dynamics of a soliton in a non-parabolic trap with the potential energy
\begin{eqnarray}
\label{U}
U\left( x\right) =\frac m2 \omega^4 x^{4}
\end{eqnarray}
(i.e. the case $r=2$). Now $R_2=\pi 2^{1/4}{\rm K}(1/\sqrt{2})\approx 0.847$, ${\rm K}(\cdot)$ being the complete elliptic integral of the first kind, and the frequency of soliton oscillations depends on the energy of the condensate [see (\ref{link_vx}) and (\ref{frequencyGP})]. The exponent defined by (\ref{x_vs_o}) is $\gamma=0.8$.

\begin{widetext}

\begin{figure}[ht]
\epsfig{file=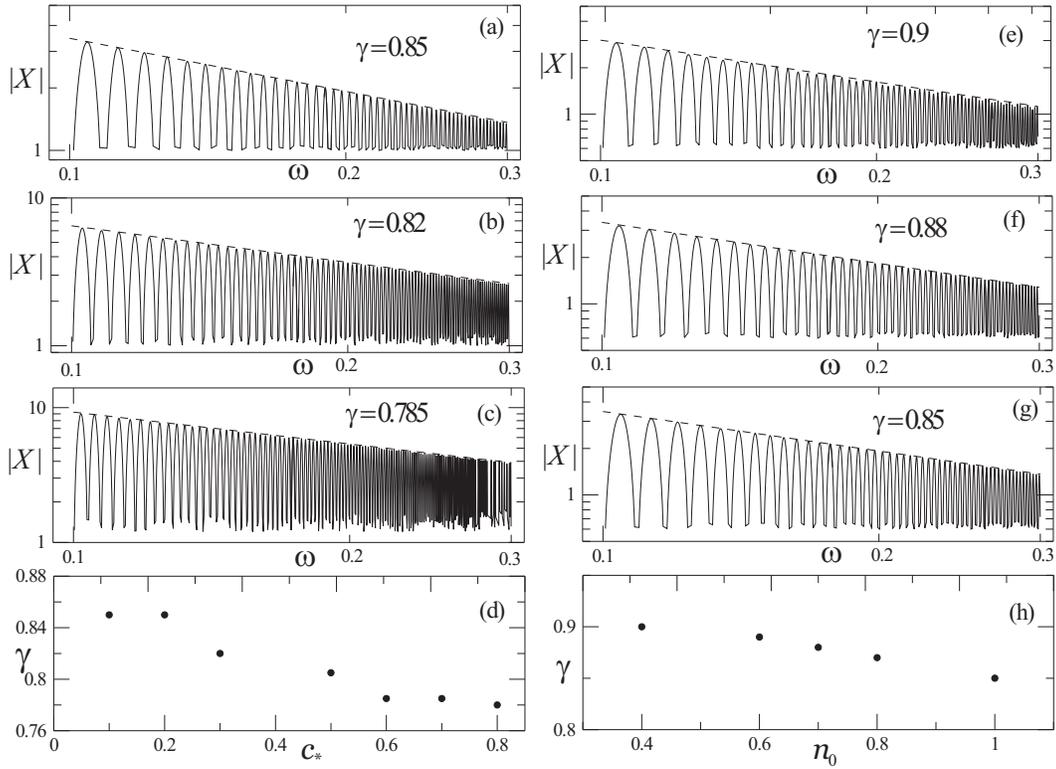,width=14cm}
\caption{Dependence of the soliton coordinate on the frequency in the logarithmic scale. Adiabatic change of the frequency is modeled by the law $\omega(t)=(1+0.001t)\cdot 0.1$. The dashed lines visualize the law $X_0=\tilde{X}_0\omega^{-\gamma}$. In panes (a), (b) and  (c) the parameters are $n_0=1$ and $v=0.1; 0.3$, and $0.6$, respectively. The matching parameter is $\tilde{X}_0=0.49; 0.98; 1.52$. In panels (e), (f) and (g) the parameters are $v=0.1$, and $n=0.4; 0.7$, and $1$, respectively. The matching  parameter is $\tilde{X}_0=0.38; 0.445; 0.49$. In panel (d) we show the dependences of  the exponent $\gamma$ {\it vs} soliton velocity $v$ at the density $n_0=1$. In panel (h) we show the dependences of  the exponent  $\gamma$ {\it vs} density $n_0$ at $v=0.1$. In the numerical calculations we used $\hbar=1$, $m=1$ and $g=1$.}
\label{fig2}
\end{figure}

\end{widetext}

The numerical study of the soliton dynamics in a  quartic trap are presented in Fig.~\ref{fig2}.
While the predicted exponential law $\omega^{-0.8}$ is now also
obtained with reasonable accuracy, there are several features
which distinguish the present case from the case shown in
Fig.~\ref{fig1}. First, one does not observe beatings
(while they are well pronounced in the case of a parabolic trap). This fact can be explained by the absence of the unique frequency of the background oscillations: in the case at hand the Ehrenfest theorem does not result in a coupled equation for the averaged coordinate of the center of mass of the condensate.  Second the dependencies of the exponent $\gamma$ of the soliton velocity and on the density appear to be decreasing functions, as it is shown in Figs.~\ref{fig2} (d) and (h).

\subsection{Dark soliton in the QNLS limit.}

As the next example we consider the  equation
\begin{eqnarray}
        i\hbar \Psi_t=
        -\frac{\hbar^2}{2m}\Psi_{xx}+
        \frac{1}{2}m\omega^2x^2\Psi+g|\Psi|^4\Psi
        -\mu\Psi\,.
\end{eqnarray}
 Now $\alpha=2$ and $r=1$. Although general approach,
similar to one developed in the preceding section is also
available in the case at hand, it becomes rather involved and
cumbersome. That is why, here we consider the physically relevant
case  of the parabolic potential which reveals the main
physical features of highly nonlinear models.

The dark soliton solution has the following form~\cite{Kolom}
\begin{eqnarray}
        &&\Psi_s(x,t) = \sqrt{n_s(x,t)}e^{i\theta_s(x,t)}
        \\
        &&n_s(x,t) = \sqrt{g}n_0-
        \frac{12\sqrt{g}n_0(c^2-v^2)e^{(x-X(t))/\ell}}{c^2\left(4+e^{(x-X(t))/\ell}\right)^2-12(c^2-v^2)}
        \nonumber \\
        &&
        \\
        &&\theta_s(x,t) =
        -\arctan\left(\frac{c^2 e^{(x-X(t))/\ell}
        -2c^2+6v^2}{6v\sqrt{c^2-v^2}}\right)
\end{eqnarray}
where $X(t)=vt+x_0$, $x_0$ is a constant, and
$\ell=\hbar/\left(2m\sqrt{c^2-v^2}\right)$.

The TF distribution now acquires the form
\begin{eqnarray}
\label{TF1}
        n_{TF}(x)=\frac{1}{\sqrt{g}}
        \sqrt{\mu-\frac 12 m\omega^2x^2}
\end{eqnarray}
and the normalization conditions defines the chemical potential
$\mu =\sqrt{2mg}\omega N/\pi$.

The energy is computed from (\ref{energy}), (\ref{density}) to be
\begin{eqnarray}
        E=\hbar \sqrt{\frac{m}{g}} \frac{\sqrt{3}}{4\sqrt{2}}
        \left(c^2-v^2\right)\ln\frac{2c+\sqrt{3}u}{2c-\sqrt{3}u}\,.
\end{eqnarray}

Taking into account that due to  (\ref{TF}) now
\begin{eqnarray}
c^2=c_0^2-\omega^2 x^2
\end{eqnarray}
and introducing the notation 
 \begin{eqnarray}  
\label{cal_E}
{\cal E}_0=\hbar n_0c_0
\end{eqnarray}
we obtain
\begin{eqnarray}
\label{energy-small}
E&=&{\cal E}_0\frac{\sqrt{3}}{4}\left(1-\frac{\omega^2x^2}{c_0^2}-\frac{v^2}{c_0^2}\right)
\nonumber\\
&\times&
\ln\frac{2\sqrt{c_0^2-\omega^2x^2}+\sqrt{3}\sqrt{c_0^2-\omega^2x^2-v^2}}{2\sqrt{c_0^2-\omega^2x^2}-\sqrt{3}
\sqrt{c_0^2-\omega^2 x^2-v^2}}\,.
\end{eqnarray}
Respectively, the energy of the zero velocity dark soliton is $E_0=E(v=0)\approx 1.14 {\cal E}_0$.

\begin{figure}[ht]
\epsfig{file=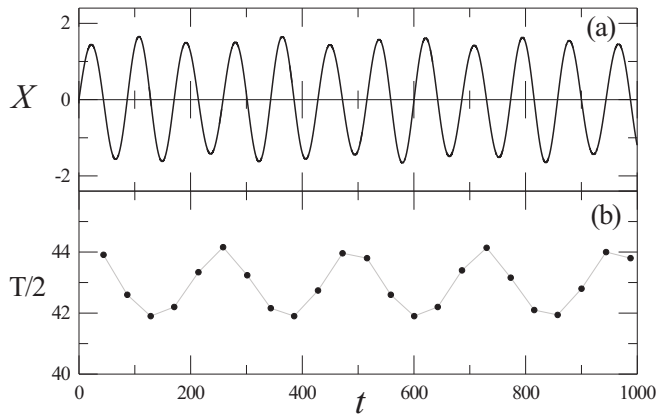,width=\columnwidth}
\caption{(a) Dependence of the position of the center of the QNLS dark soliton on time for   $\omega=0.1$, $n_0=1$, and $v=0.1$ (b) Time dependence of the half-period $T/2$ substracted from figure (a).
As before we take $\hbar=1$, $m=1$ and $g=1$.}
\label{figfour}
\end{figure}

In Fig.~\ref{figfour}(a) we present a typical trajectory of the QNLS dark soliton in a constant trap. One of the main features observed is that the dynamics is not strictly periodic, but undergoes slow modulations (see Fig.~\ref{figfour}). The averaged frequency of the dynamics shown is approximately 0.07 (this corresponds to the relation $\omega_s\approx 0.7\omega$) while the frequency of the large oscillations of the period is approximately 5 times less. It is worth pointing out that the theoretical prediction for the frequency of the large amplitude (slow) dark solitons in the QNLS model gives $\omega_s\approx 0.6572\omega$ (see the Table~\ref{tab:SolitonCharacteristics} below) while small amplitude solitons should oscillate  with the frequency close to $\omega_s\approx \omega$ (see the discussion in Sec.~\ref{sec:smallamp}).

\begin{widetext}

\begin{figure}[ht]
\epsfig{file=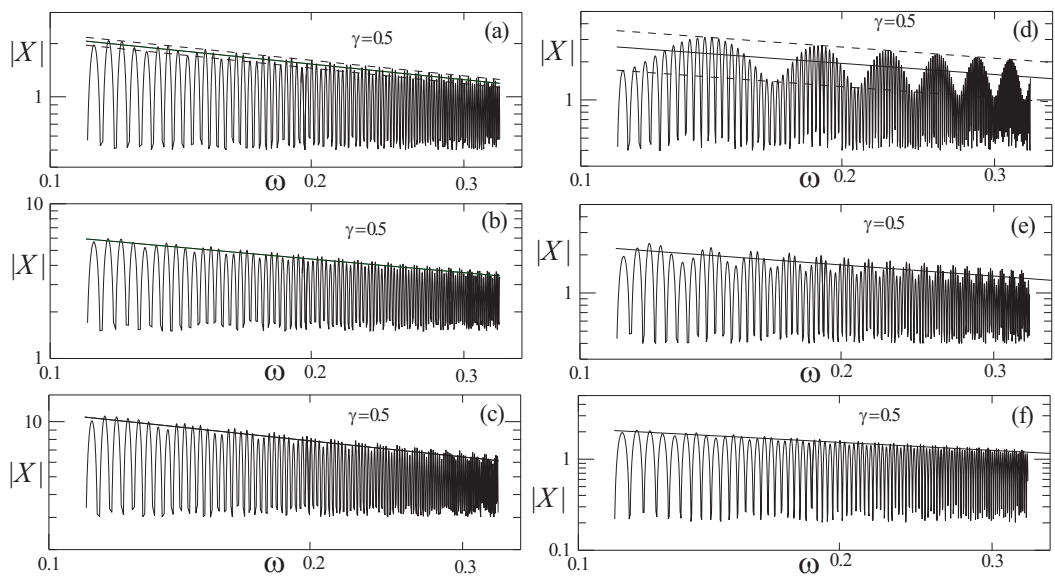,width=14cm}
\caption{Dependence of the soliton coordinate on the frequency adiabatically varying according to the law $\omega(t)=(1+0.001t)\cdot 0.11$. Straight lines show the law $X_0\propto \omega^{-\gamma}$. In the panels (a) and (d) we show also the laws $X_{0,\pm}\propto \omega^{-\gamma_\pm}$ [see (\ref{pm})] by dashed lines.
In figures (a), (b) and (c) parameters are $n_0=1$, and $v=0.14; 0.42$, and $0.85$, correspondingly.
In figures (d), (e) and (f) parameters are $v=0.14$, and $n_0=0.4; 0.6$, and $1$, correspondingly.
In the numerical calculations we take $\hbar=1$, $m=1$ and $g=1$.}
\label{fig3}
\end{figure}
\end{widetext}

For the next step we studied the adiabatic dynamics of the QNLS dark soliton in a slowly varying trap. The respective results are shown in Fig.~\ref{fig3}.

Like in the case of the NLS dark soliton in a parabolic trap one can observe beating of the solution. From the left column [panels (a) to (c)] one detects increase of the frequency with increase of the initial value of the velocity, what is expectable in view of the above discussion. The right column [panels (d) to (f)] show that the frequency of soliton oscillations decay when the background density increases. This last fact is also explained in view of the above discussion, by the fact that increase of the local density subject to constant velocity $v$ results in increase of the relativistic factor $c^2-v^2$, and thus in bigger difference between the speed of sound and soliton velocity. In all the cases however one observes well pronounced scaling law with the exponent $\gamma=0.5$ for relatively large range of the parameters.



\section{Limiting cases}

\subsection{Small velocity solitons.}

\subsubsection{General relations.}

As we have seen above, increase of the
power of the nonlinearity (i.e. of the exponent $\alpha$)  makes the problem of the computing the frequency and dependence of the frequency on the amplitude of oscillations rather complicate, not allowing one to obtain a general formula linking $X_0$ and $\omega$ for arbitrary $\alpha$. It turns out however, that the problem can be solved in the limit of small velocities: $v\ll c$. To this end we take into account that the static dark soliton for any $\alpha>0$ has zero amplitude in its center, and hence the limit of small velocities corresponds to the limit of small $X$. Then, expanding Eq. (\ref{din_eq_1}) with respect to $v^2$ and $X^2$, one obtains in the leading orders
\begin{eqnarray}
\label{en_expan}
E=E_0+ \frac{\partial E_0}{\partial v^2}v^2+ \frac{\partial E_0}{\partial X^2}X^2
\end{eqnarray}
where the subindex ``$0$'' stands to indicate that the respective quantities are computed in the point $X=0$ and $v=0$. This formula must be viewed as a standard expression
\begin{eqnarray}
\label{Es}
        E_s=\frac{m_s}{2}\left(v^2+\omega_s^2x^2\right)
\end{eqnarray}
for the enrgy of a harmonic oscillator having mass $m_s$ and  frequency $\omega_s$. Comparison of (\ref{en_expan}) with (\ref{Es}) gives the expressions for the effective mass
\begin{eqnarray}
        \label{eff_mass}
        m_s=2\frac{\partial E_0}{\partial v^2}
\end{eqnarray}
and for the frequency of oscillations
\begin{eqnarray}
        \label{eff_freq}
        \omega_s=\left(\frac{\partial E_0}{\partial X^2}\Bigg/\frac{\partial E_0}{\partial v^2} \right)^{1/2}
\end{eqnarray}
of a small amplitude dark soliton.

Thus to compute frequency dependence of the amplitude of the soliton oscillations from Eq.~(\ref{NLS}) we have to expand the energy $E(c(X),v)$ for small $X$ and $v$.
It is convenient to do this in dimensionless variables which we define as follows:
\begin{eqnarray}
\label{dimless_var}
\psi =n_0^{-1/2}\Psi,\quad \zeta=\frac{mc_0}{\hbar }
\sqrt{\frac{2}{\alpha } }x,\quad \tau=\frac{c_0^2m}{\hbar\alpha}t\,,
\end{eqnarray}
allowing one to rewrite (\ref{NLS}) in the dimensionless form
\begin{eqnarray}
\label{NLS-dimless}
i\psi_\tau=-\psi_{\zeta\zeta}+ (|\psi|^{2\alpha}-1)\psi\,	
\end{eqnarray}
[here we used the relations (\ref{speed})].
Also we will use the notation $V=\sqrt{\frac{\alpha}{2}}\frac{v}{c_0}$. 
Then looking for the dark soliton solution, i.e. one having form (\ref{running}) and thus depending only on the running variable $x-vt$ (what in dimensionless variables means dependence on $\zeta-2V\tau$), and representing $\psi=\eta\exp(i\theta)$ one obtains (see Appendix~\ref{link}) the link
\begin{eqnarray}
\label{theta}
 \theta_{\zeta }=-V\frac{ 1-\eta^{2}}{\eta^{2}}   
\end{eqnarray}
and the equation for $\eta$ (notice that according to (\ref{bound}) the boundary conditions now are  $\eta\rightarrow 1$ and $\theta_\zeta \rightarrow 0$ as $\zeta \rightarrow \pm \infty$)
\begin{eqnarray}
\eta_{\zeta \zeta }+\left( 1-\eta^{2\alpha }\right) \eta-V^{2}\frac{1-\eta^{4}}{\eta^{3}}=0.
\label{ddr}
\end{eqnarray}
The last equation can be integrated with respect to $\zeta$ what gives
\begin{eqnarray}
P^{2}=\frac{1}{\alpha +1}\left( \eta^{2\alpha +2}-1\right) +1-
\eta^{2} -V^{2}\frac{\left( 1-\eta^{2}\right) ^{2}}{\eta^{2}}
\label{p1}
\end{eqnarray}
where we designated $\eta_\zeta=P\equiv P(\eta)$.

Now the energy of the soliton can be rewritten in the form (see Appendix~\ref{ap:energy})
\begin{eqnarray}
\label{energy_v}
E={\cal E}_0\frac{2\sqrt{2}}{\sqrt{\alpha}}\int\limits_{\eta_{m}}^{1}\left[ P^{2}(\eta)+\frac{\left( 1-\eta^{2}\right) ^{2}}{
\eta^{2}}V^{2}\right] \frac{d\eta}{P(\eta)}  
\label{Er1}
\end{eqnarray}
where ${\cal E}_0$ was introduced in (\ref{cal_E})
and $\eta_m$ determines the soliton amplitude in its center and solves the equation
\begin{eqnarray}
\label{P0}
        P(\eta_m)=0.
\end{eqnarray}

For a particular case of the zero-velocity dark soliton one has
\begin{eqnarray}
E_{0}={\cal E}_0G_\alpha,
\label{Ed}
\end{eqnarray}
where
\begin{eqnarray*}
G_\alpha=\frac{2\sqrt{2}}{\sqrt{\alpha(\alpha +1)}}\int\limits_{0}^{1}\sqrt{\eta^{2\alpha +2}-\left(
\alpha +1\right) \eta^{2}+\alpha }d\eta\,.
\end{eqnarray*}
(Notice that  in this case $\eta_{m}=0$,  but
$P(\eta=0)=\sqrt{\alpha/(\alpha +1)}\neq 0.$) 
Particular values of the energy for some relevant models are presented in Table~\ref{tab:SolitonCharacteristics}.

\subsubsection{Effective mass of a dark soliton.}

Let us consider now a soliton moving with a small velocity, $V \ll 1$.
To execute the expansion of the energy we first notice that from (\ref{P0}) and (\ref{p1}) 
it follows that in the leading order
\begin{eqnarray}
\eta_{m}\approx \sqrt{\frac{\alpha +1}{\alpha }}V.  \label{rm}
\end{eqnarray}
Next, we introduce a constant $\eta_0$ which satisfies the condition $\eta_m\ll\eta_0\ll 1$ and split the integral in (\ref{energy_v}) in two ones: $E=E_1+E_2$ where
\begin{eqnarray*}
E_1= {\cal E}_0\frac{2\sqrt{2}}{\sqrt{\alpha}}\int\limits_{\eta_0}^{1}[\cdots]\frac{d\eta}{P(\eta)},\quad E_2={\cal E}_0\frac{2\sqrt{2}}{\sqrt{\alpha}}
\int\limits_{\eta_m}^{\eta_0}[\cdots]\frac{d\eta}{P(\eta)}.
\end{eqnarray*}
As it follows from (\ref{p1}),
\begin{eqnarray*}
\frac{\partial P}{\partial V^{2}}=-\frac{1}{2P}\frac{\left( 1-\eta^{2}\right) ^{2}}{\eta^{2}}
\end{eqnarray*}
and thus in the limit $V\to 0$
\begin{eqnarray}
\label{dE1}
\frac{dE_{1}}{dV^{2}}={\cal E}_0\frac{2\sqrt{2}}{\sqrt{\alpha}}\int\limits_{\eta_0 }^{1}\left[ \frac{1}{\alpha +1}
\left( \eta^{2\alpha +2}-1\right)+1 -\eta^{2} \right] 
\nonumber \\ \times 
\frac{\partial}{\partial V^{2}
}\frac{1}{P}d\eta
\approx
{\cal E}_0\frac{\sqrt{2}}{\sqrt{\alpha}}\int\limits_{\eta_0 }^{1}\frac{
\left( 1-\eta^{2}\right) ^{2}}{P_0(\eta)}\frac{d\eta}{\eta^{2}}
\nonumber \\
\approx {\cal E}_0\frac{\sqrt{2}}{\sqrt{\alpha}}\int\limits_{0}^{1}\frac {d}{d\eta}\left(\frac{
\left( 1-\eta^{2}\right) ^{2}}{P_0(\eta)}\right)\frac{d\eta}{\eta}+{\cal E}_0\frac{\sqrt{ 2(1+\alpha)}}{\alpha\eta_0}
\end{eqnarray}
(to obtain the last line we substituted the lower limit by zero, due to fast convergence of the integral, and integrated by parts).

To calculate the derivative of $E_{2}$ we take into account that $\eta$ is small over the whole range of integration. Thus
\begin{eqnarray*}
\label{E2}
E_{2}&\approx& {\cal E}_0\frac{2\sqrt{2}}{\sqrt{\alpha +1}}\int\limits_{\eta_{m}}^{\eta_0 }
\frac{\eta}{\sqrt{\eta^{2}-\eta_{m}^{2}}}d\eta
\nonumber \\
&\approx& {\cal E}_0 \frac{2\sqrt{2}}{\sqrt{\alpha +1}}\eta_0 -{\cal E}_0\sqrt{
\frac{2 }{\alpha +1}}\frac{\eta_{m}^{2}}{\eta_0 }
\end{eqnarray*}
and in the limit $V\to 0$ [due to (\ref{rm})],
\begin{eqnarray}
\frac{dE_{2}}{dV^{2}}=-{\cal E}_0\frac{\sqrt{2(\alpha +1)}}{\alpha\eta_0 }
.  \label{dE2}
\end{eqnarray}
Sum of (\ref{dE1}) and (\ref{dE2}) gives us the derivative we are looking for:
\begin{eqnarray}
\label{dE}
\frac{dE_0}{dV^{2}}={\cal E}_0F_\alpha,\quad F_\alpha=\sqrt{\frac{2}{\alpha}} \int\limits_{0}^{1}\frac{1}{\eta}\frac{
d}{d\eta}\frac{\left( 1-\eta^{2}\right) ^{2}}{P_0(\eta)}d\eta\,.
\end{eqnarray}
(We recall that the subindex ``$0$'' on the l.h.s. stands for $v=0$ and $X=0$.)
Finally, the definition of the effective mass (\ref{eff_mass}), the explicit expression for the momentum $P_0$ (\ref{p}), and the last formula (\ref{dE}) yields the general expression for the mass of the dark soliton:
\begin{eqnarray}
\label{EM}
m_s=\alpha F_\alpha \frac{{\cal E}_0}{c_0^2}\,.
\end{eqnarray}

In order to relate the mass of the soliton to the atomic mass $m$, we recall that $-dE/d\mu=N$ where $N$ is the negative ``total number of particles'' associated with the soliton. Thus, $m_*=m_s/N$ can be considered as the effective mass of a ``solitonic'' particle. In the limit $V\to0$ the above  quantities can be easily calculated to give
\begin{eqnarray}
        N=\frac{{\cal E}_0}{m c_0^2} N_\alpha,\quad N_\alpha=\sqrt{2\alpha}\int_{0}^{1}(1-\eta^2)\frac{d\eta}{P_0(\eta)}
\end{eqnarray}
and
\begin{eqnarray}
        m_{\ast}=\frac{\alpha F_\alpha}{N_\alpha}m\,.
\end{eqnarray}

In Table~\ref{tab:SolitonCharacteristics} we present examples of the effective mass for two relevant cases. From the provided values one can see that the effective mass of the soliton particle is bigger than the mass of a free particle:  $m_\ast>m$.

\subsubsection{Frequency of oscillations of a dark soliton.}

To conclude this subsection we compute the frequencies of oscillations of dark solitons in a trap, what can be done using the relation (\ref{eff_freq}). To this end we notice that including the trap potential into the scheme developed in the preceding subsection, can be done by simple change the chemical potential $\mu$ by $\mu-U(X)$. Thus for the parabolic trap we have
\begin{eqnarray}
        \frac{\partial E_0}{\partial X^2}=-\frac{\omega^2}{2}\frac{\partial E_0}{\partial \mu}=\frac{\omega^2N}{2}
\end{eqnarray}
This leads us to the formula
\begin{eqnarray}
        \omega_s=\frac{N_\alpha}{\alpha F_\alpha}\omega\,.
\end{eqnarray}

\begin{widetext}

\begin{table}[t]
        \begin{center}
                \begin{tabular}{c||c|c|c|c}
                $\alpha$        &  $E_0$ &  $|N|$ & $m_*$& $\omega_s$  \\
                        \hline\hline
                        $2$   & $\displaystyle{2\sqrt{3}\ln\frac{1+\sqrt{3}}{\sqrt{2}}\,{\cal E}_0}$ &    $\displaystyle{4\sqrt{3}\ln\frac{1+\sqrt{3}}{\sqrt{2}}\,\frac{{\cal E}_0}{m c_0^2}}$& $\displaystyle{\frac{\sqrt{3}+2\ln\frac{1+\sqrt{3}}{\sqrt{2}}}{2\ln\frac{1+\sqrt{3}}{\sqrt{2}}}\,m}$ & 0.6572$\,\omega$ \\
                        $1$   &  $\displaystyle{\frac{4\sqrt{2}}{3}\;{\cal E}_0}$ &  $\displaystyle{2\sqrt{2}\,\frac{{\cal E}_0}{m c_0^2}}$
                        & $2m$ & $\displaystyle{\frac{\omega}{\sqrt{2}}}$ \\
                \end{tabular}
        \end{center}
        \caption{Characteristics (the zero-velocity energy $E_0$, the number of particles $N$, the effective mass $m_{\ast}$, and the frequency of oscillations in the parabolic trap $\omega_s$) of dark solitons with small velocities for different powers of the nonlinearity $\alpha$.}
        \label{tab:SolitonCharacteristics}
\end{table}

\end{widetext}

\subsection{Analysis based on the perturbation theory.}
\label{sec:perturbed}

In Ref.~\cite{KP} it has been argued that the phenomenological approach formulated above can be justified from the viewpoint of the original GP equation with help of the perturbation theory for dark solitons~\cite{KV94,BK1} when motion occurs in a constant parabolic trap. The proof was based on a possibility of effective factorization of the solution on the constant background and the dark soliton solution moving against it. In the case of time dependent trap, the background cannot be considered as a constant, and the theory requires revision. The goal of the present subsection is to develop the modification of the perturbation theory and to obtain from it the exponent $\gamma$ which describes change of the amplitude of oscillations of the GP dark soliton in a parabolic trap.

To this end we start with the dimensionless form of the one-dimensional GP equation [the variable are introduced in (\ref{dimless_var}), see also Eq. (\ref{NLS-dimless})]
\begin{eqnarray}
\label{NLS_new}
        i\psi_\tau+\psi_{\zeta\zeta}-\frac12\nu^2\zeta^2\psi-|\psi|^2\psi=0\,,
\end{eqnarray}
where $\nu \equiv \nu \left( \tau\right) =\hbar /\left(
2^{1/2}c_{0}^{2}m\right) \omega \left( t\right) .$ We assume that $\omega
\left( t\right) $ is a slow function of time, what is expressed by   
the adiabaticity condition $\frac{1}{\omega^2}\left|\frac{d\omega}{d t}\right|\ll 1.$ Accordingly, 
$\nu \left( \tau \right) $ is also a slow function. Notice that $\nu \left(
0\right) =\nu _{0}\ll 1$ as a condition for the local density approximation.
We look for a solution of (\ref{NLS_new}) in a form of the ansatz (analogous of the well known lens transformation)
\begin{eqnarray}
\label{ansatz}
        \psi(\zeta,\tau)=
        e^{-if(\tau)\zeta^2}\frac{1}{\sqrt{L(\tau)}}\phi(\xi,\tilde{\tau}(\tau)),
\end{eqnarray}
where $\xi$ is a function on time and on spatial coordinate given by $\xi=\zeta/\sqrt{L(\tau)}$, while $\ttau$ is a new temporal variable related to the old one by the equation $\ttau_\tau=1/L$.   The functions $L(\tau)$ and $f(\tau)$ are to be determined below. Substitution of (\ref{ansatz}) into (\ref{NLS_new}) yields
\begin{eqnarray}
\label{eq1}
        i\phi_\tau+\phi_{\xi\xi}-|\phi|^2\phi-\left(f_\tau+4f^2+\frac 12\nu^2\right)L^2\xi^2\phi
        \nonumber \\
        -\frac{i}{2} \left(L_\tau-4fL\right)\phi
        -\frac{i}{2} \left(L_\tau-8fL\right)\xi\phi_\xi=0\,.
\end{eqnarray}
Let us now require the trap frequency of the new equation (i.e. term proportional to $\xi^2\phi$) to be constant, say $1/2 \nu_0^2$, and dissipative terms, i.e. the linear with respect to $\phi$, to vanish. This gives us two equations: 
\begin{eqnarray}
\label{eq_f}
        \left(f_\tau+4f^2+\frac12\nu^2\right)L^2=\frac 12\nu_0^2
\end{eqnarray}
and
\begin{eqnarray}        
        \label{eq_ell}
         L_\tau=4fL\,.
\end{eqnarray}
The obtained equations will be supplied by the natural initial conditions $f(0)=0$ and $L(0)=1$. Then Eq. (\ref{eq1}) takes the form
\begin{eqnarray}
\label{GP2}
        i\phi_\tau+\phi_{\xi\xi}-|\phi|^2\phi -\frac 12 \nu_0^2\xi^2\phi =-2ifL\xi\phi_\xi\,.
\end{eqnarray}
We emphasize that the last equation is {\em exact} with no approximation made, so far.

Before the analysis of (\ref{GP2}), let us consider Eqs. (\ref{eq_f}) and (\ref{eq_ell}) in more details. They can be reduced to a single equation for $L$:
\begin{eqnarray}
\label{LL}
\frac{L}{2}L_{\tau\tau} +\nu^2L^2-\nu_0^2=0,
\end{eqnarray}
Due to adiabaticity the first term in (\ref{LL}) is small in comparison with
the other ones. Neglecting that  term, we find in the leading order
\begin{eqnarray}
L=\frac{1}{\tilde{\tau}_{\tau }}=\frac{\nu _{0}}{\nu \left( \tau \right) }\quad
\mathrm{and} \quad f=-\frac{\nu _{\tau }}{4\nu }.
\end{eqnarray}
Then, simple estimates give $L_{\tau \tau }\sim \nu ^{2}\left( 
\frac{1}{\omega ^{2}}\left| \frac{d\omega }{dt}\right| \right) ^{2}\ll \nu ^{2}$ and $L \approx \nu_{0}/\nu \sim 1$, what justifies the approximation made.

Next we introduce the notation $R$ for the r.h.s. of (\ref{GP2}): $R\equiv-2ifL\xi\phi_\xi$.  Since $f$ is small (because of the adiabaticity of the change of $\nu$) this term gives us an perturbation, which is a complementary to the perturbation introduced by a constant parabolic trap, provided $\nu_0$ is small, the case considered in detail in \cite{BK1}. Due to their smallness, the effect of different perturbations on the dynamics of the soliton center is additive, allowing one to compute only the contribution of $R$ to the dynamical equation of the soliton center and add it to the equation describing soliton in a stationary potential obtained in~\cite{BK1} [see Eq.~(32) there]. We skip description of tedious but straightforward calculations~\footnote{Derivation of the final equation is reduced to substitution of the expression for the unperturbed dark soliton (\ref{GPsoliton}) into the function $R$ with subsequent calculation of the equations of the direct perturbation theory for dark solitons given by (B1)-(B3) in Ref.~\cite{BK1}, where for our purposes it is enough to keep only the leading order.} and present only the final result:   
the equation for the soliton coordinate, in terms of the rescaled by $L$ variables, is given by
\begin{eqnarray}
 \frac{d X}{d\tau}=V-\frac 12 \nu_0^2\int\limits_0^\tau  X(\tau^\prime)d\tau^\prime-\frac{1}{4}\left(\frac{\nu_0}{\nu}\right)_\tau V X\,.
\end{eqnarray}

Next we differentiate the last equation with respect to $\tau$  and eliminate the ``dissipative'' term by means of the substitution
\begin{eqnarray}
  X=Y(\ttau)e^{\vartheta(\ttau)};
\end{eqnarray}
where
\begin{eqnarray}
 \vartheta= -\ln\left(\frac{\nu}{\nu_0}\right)^\delta,\qquad
 \delta=- \frac{V}{8}\,.
\end{eqnarray}
Having done this and restoring the original variables we arrive at the final formula  
\begin{eqnarray}
        X_0\propto\nu^{-\gamma},\qquad  \gamma=\frac 12+\delta\,.
\end{eqnarray}

Comparing this result with (\ref{x_v_om_harm1}) and (\ref{x_v_om_harm2}) one can see a remarkable agreement. The perturbation theory, valid for relatively low densities of the condensate, and thus to the Gaussian background, corrects the law (\ref{x_v_om_harm2}) based on the phenomenological approach, by means of small shift (recall that $V\ll 1$ and thus $\delta\ll 1$) toward larger exponent which in the TF limit is given by (\ref{x_v_om_harm1}). Moreover, the perturbation theory introduces an explicit dependence of the exponent on the velocity (at this point it is relevant to recall also that the frequency itself does not depend on the soliton velocity). Finally we mention that the obtained result corroborates with the numerical results on the dependence of the exponent $\gamma$ on the soliton velocity and on the density of the background depicted in Fig.~\ref{fig1} (d) and (h).

\subsection{A comment on small-amplitude solitons.}
\label{sec:smallamp}

Small amplitude dark solitons of the NLS equation with a polynomial nonlinearity of any power of the nonlinearity are described by the KdV equation~\cite{BKP} and they move with the sound velocity (or more precisely with a velocity slightly deviating from the sound velocity). While self consistent reduction of the 3D GP equation to one-dimensional KdV equation seems to be not possible for realistic condensates (as it is explained in Ref.~\cite{BK2}), the KdV being rather academic than practical allows one to predict some features of the underline GP equation.

Moreover, one can easily argue, that the small-amplitude limit of a dark soltion in a parabolic trap is not available. Indeed, existence of a soliton in a trap implies smallness of the soliton width $\ell$ compared to the trap width $\sqrt{\hbar/(m\omega)}$. Using the expression for the width of a dark soliton, which is given by (\ref{GPsoliton}), one immediately obtains the limitation $ \hbar\omega/m\lesssim c^2-v^2$. Thus the existence of a trap does not allow the truth small amplitude limit, which would correspond to $v\to c$.

 Let now {\it formally} compute the half-period of oscillations of a small-amplitude soliton in a parabolic trap. Under the half-period we understand the time necessary for a soliton to pass the distance between two turning points.   To this end associate the velocity
\begin{eqnarray}
        c(x)= \omega\sqrt{\frac{\alpha}{2}(x_0^2-x^2)} 
\end{eqnarray}
where $x_0=2\mu/(m\omega^2)$, with the velocity of the soliton. Then direct computation gives
\begin{eqnarray}
\label{om_small_amp}
        \omega_{sol}=\frac{\pi}{\int_{-x_0}^{x_0}\frac{dx}{c(x)}}
        =\sqrt{\frac{\alpha}{2}}\omega\,.
\end{eqnarray}
Thus for the small amplitude GP dark soliton in a parabolic trap we obtain $\omega/\sqrt{2}$, what coincides with the results known for relatively large velocity for the soliton.  Eq. (\ref{om_small_amp}) also gives $\omega_s=\omega$ for $\alpha=2$, the result recently reported in \cite{K}.

We emphasize however that presently there are no available results confirming validity of the law  (\ref{om_small_amp}) for small amplitude NLS solitons. The main physical reason for this, mentioned in~\cite{BK1}, is that in the vicinity of the turning points the density becomes small enough making the problem to be linear and thus not allowing solitonic propagation due to dominating dispersion. Mathematically, the problem occurs due to divergence (see e.g. the second of equations (11) in Ref.~\cite{K}) of the small amplitude expansion near the points where the condensate density, and thus the speed of sound, in the TF approximation becomes zero (see also discussion of the failure of the small amplitude limit in~\cite{PFK}).

\section{Conclusion}

In the paper we presented development of the theory suggested in the earlier publication~\cite{KP}, providing detailed description of the one-dimensional dynamics of a dark soliton in a Bose-Einstein condensate confined by an external potential. The theory is based on the local density approximation and allows one to interpret the dark soliton as a hamiltonian particle.
We addressed various generalizations of the theory including the nonlinearity of a general polynomial type as well as non-parabolic potential.  We have obtained that the dependence of the amplitude of the soliton oscillations in a external trap depends on the adiabatically changing frequency through the scaling law $X_0\propto \omega^{-\gamma}$ where the exponent $\gamma$ depends on the type of the nonlinearity and on the type of the confining potential. It turns out also that the frequency dependence of the amplitude of the oscillations depends also on the density of the condensate and on the initial velocity, even in the cases when the frequency itself is independent on the above quantities as in the case of the standard nonlinear Schr\"{o}dinger dark solitons. Also the obtained scaling law in a general case appears to be very different from the corresponding law for the linear oscillator. 

We dedicated special attention to the cases of dark solitons within the framework of the Gross-Pitaevskii and quintic nonlinear Schr\"{o}dinger models. We also have shown that in the limiting case of slow, and thus large-amplitude, solitons one can obtain the general explicit expressions for the effective mass of the dark soliton, considered as a quasi-particle, and for the frequency of its oscillations in the external confining trap.

The results have been verified numerically, showing good agreement with theory, and were shown to be in agreement with outcomes of the direct perturbation theory for solitons.

\acknowledgements

V.V.K. thanks D. E. Pelinovsky for sending the results of Ref~\cite{PFK} and J. Brand for the useful comments. V.A.B. acknowledges support of the FCT grant SFRH/BPD/5632/2001.
The work of V. A. B. and V. V. K. was supported by the grant POCI/FIS/56237/2004.


\appendix

\section{Estimates for the background}
\label{app:F}

In the present appendix we provide estimates for the last two terms in (\ref{energy_full}). For the sake of simplicity the consideration will be restricted to the case of a polynomial parabolic trap (\ref{plynom}).

Let us consider the behavior of the function $F(x)$ in the vicinity of the point $x\ll L_0$ (we recall that $L_0$ is an effective trap length). The background is obviously an even function of the trap what allows us to look for its solution in a form of the expansion
\begin{eqnarray}
F=1+\sum_{k=1}^{\infty} F_k\zeta^k,\qquad \zeta=x^2\, .
\end{eqnarray}
More specifically we are looking for the coefficients $F_k$, all of which become zero in the homogeneous case when $\omega=0$. It follows directly from (\ref{eq:F}) that
\begin{eqnarray}
        \mu=gn_0^\alpha-\frac{\hbar^2}{m}F_1\, .
\end{eqnarray}
In the homogeneous condensate the chemical potential is given by $\mu_0=gn_0^\alpha$ and thus there should be verified that $F_1\ll mgn_0^\alpha/\hbar^2$ for $\omega$ small enough.

Next from (\ref{eq:F}) one can obtain the recurrent formulas
\begin{eqnarray}
        \label{recur1}
        \frac{\hbar^2}{2m}(2k+2)(2k+1)F_{k+1}=
         \frac{gn_0^{\alpha}}{k!} \left(\frac{d^kF^{2\alpha+1}}{d\zeta^k} \right)_{\zeta=0}
                \nonumber \\
        -\mu F_k\qquad \mbox{for $k<r$}
        \\
        \label{recur2}
        \frac{\hbar^2}{2m}(2r+2)(2r+1)F_{r+1}=
         \frac{gn_0^{\alpha}}{r!} \left(\frac{d^rF^{2\alpha+1}}{d\zeta^r} \right)_{\zeta=0}
                \nonumber \\
        -\mu F_r+ \frac{m}{2}\omega^{2r}\,.
\end{eqnarray}
In order to satisfy the constrain $F_k=0$  at $\omega=0$, we require $F_{r+1}\ll F_r$. From (\ref{recur1}) and (\ref{recur2}) we obtain the following asymptotic relations
\begin{eqnarray*}
        &&F_k={\cal O}\left(F_r\right)={\cal O}\left(\omega^{2r}\right),\qquad k\leq r
        \\
        &&F_{r+1}=o\left(\omega^{2r}\right)
\end{eqnarray*}
which in their turn guarantee the smallness of the integrals
\begin{eqnarray*}
\int\limits_{|x-X|<\delta}\frac{F_{xx}n(x)}{F}dx
\propto\omega^{2r}
\,\,\,\mbox{and}\,\,\,
 \int\limits_{|x-X|
 <\delta}\frac{d^2n_0(x)}{dx^2}dx\propto\omega^{2r}
\end{eqnarray*}
when $\omega\to 0$.

\section{On the link among formulas (\ref{NLS-dimless}), (\ref{theta}) and (\ref{ddr}) }
\label{link}

 In terms of the amplitude $\eta$ and the phase $\theta$, both depending on $\zeta-2V\tau$, Eq. (\ref{NLS-dimless}) can be rewritten in the form of a system
\begin{eqnarray}
\label{link1a}
	&&-2V\theta_\zeta=\frac{\eta_{\zeta\zeta}}{\eta}-\theta_\zeta^2+1-\eta^{2\alpha}
	\\
	\label{link1b}
	&&2V\eta_\zeta=2\eta_\zeta\theta_\zeta+\eta\theta_{\zeta\zeta}
\end{eqnarray}
Multiplying (\ref{link1b}) by $\eta$, integrating with respect to $\theta$ and using the boubdary conditions $\eta\to 1$ and $\theta\to$const as $\zeta\to\pm\infty$, one obtains the link (\ref{theta}).

In order to obtain (\ref{ddr}) it is enough to substitute $\theta_\zeta$ expressed in terms of $\eta$ through the relation (\ref{theta}) in  (\ref{link1a}) and multiply the result by $\eta$.

\section{Adiabatic integral for the GP soliton
in a polynomial trap}
\label{ap:adint}

The adiabatic integral for the GP dark soliton is computed,
using (\ref{AIint}) and links (\ref{link_vx}) and (\ref{speed})
for $\alpha=1$, as follows
\begin{widetext}
\begin{eqnarray}
        \label{AI_GP1}
        I&=&-4\hbar\int_{0}^{X_0}n\left(\frac vc \sqrt{1-\frac{v^2}{c^2}}+
        \arcsin\left(\frac vc\right)\right)dx
        \nonumber \\
        &=&
        \frac{2^{2+1/(2r)}m^{1-1/(2r)}\hbar}{r g \omega}
        \int_{0}^{c_*}\frac{v(v^2+u^2)}{\left(u^2-v^2\right)^{(2r-1)/(2r)}}
                \left(\frac {vu}{v^2+u^2}+
        \arcsin\left(\frac {v}{\sqrt{v^2+u^2}}\right)\right)dv
        \nonumber \\
        &=& \frac{\hbar m^{1-1/(2r)}u^{2+1/r}}{g \omega}G_r
\end{eqnarray}
where the constant $G_r$ is given by
\begin{eqnarray}
        \label{Gr}
        G_r=\frac{2^{2+1/(2r)}}{r}\int_{0}^{1}\frac{y(1+y^2)} {\left(1-y^2\right)^{1-1/(2r)}}\left[
        \frac{y}{1+y^2}+\arcsin\left(\frac{y}{\sqrt{1+y^2}}\right)
        \right]dy\,.
\end{eqnarray}

\end{widetext}

\section{Calculation of the energy (\ref{energy_v})}
\label{ap:energy}

Starting with the definition (\ref{energy}), (\ref{density}), written as
\begin{eqnarray}
E=E_0\int_{-\infty}^{\infty} \Bigg[ P^{2}+\frac{1}{\alpha +1}\left( \eta^{2\alpha +2}-1\right)+1
-\eta^{2}
\nonumber \\
+\frac{\left( 1-\eta^{2}\right) ^{2}}{\eta^{2}}V^{2}\Bigg]
d\zeta  \label{EA}
\end{eqnarray}
and excluding $P(\eta)$ with the help of $\left( \ref{p}\right) $ one obtains
\begin{eqnarray*}
E&=&2E_0\int_{-\infty }^{\infty }\left[ \frac{1}{\alpha +1}\left( \eta^{2\alpha
+2}-1\right) +1-\eta^{2}\right] d\zeta
\\
&=&4E_0\int\limits_{\eta_{m}}^{1}\left[ \frac{1}{\alpha +1}\left( \eta^{2\alpha
+2}-1\right) +1-\eta^{2} \right] \frac{d\eta}{P(\eta)}\,.  \label{Er}
\end{eqnarray*}
Formula (\ref{energy_v}) follows from the last equality.


\begin{thebibliography}{99}

\bibitem{soliton} S. P. Novikov, S. V. Manakov, L. P. Pitaevskii, and V. E. Zakharov, {\em Theory of solitons: Inverse Scattering Method} (Consultants Bureau, New York, 1980).

\bibitem{kinetic} V. E. Zakharov, Sov. Phys. JETP {\bf 33}, 538 (1971).

\bibitem{Pit1} L.~P.~Pitaevskii, and S.~Stringari, {\em
  Bose-Einstein Condensation} (Oxford University Press, Oxford,
  2003).
\bibitem{termin} According to widespread terminology a dark (or grey) soliton is
a localized decrease of density, propagating in a homogeneous medium without
deformation of its form.

\bibitem{Tsuzuki} T.~Tsuzuki, J. Low Temp. Phys. {\bf 4}, 441 (1971).

\bibitem{experim} M.~R.~Andrews,
 D.~M.~Kurn, H.-J.~Miesner,
D.~S.~Durfee, C.~G. Townsend, S.~Inouye, and W.~Ketterle,
Phys. Rev. Lett. {\bf 79}, 553 (1997);
 S.~Burger, K.~Bongs, S.~Dettmer, W.~Ertmer,  K.~Sengstock,
A.~Sanpera, G.~V.~Shlyapnikov, and M.~Lewenstein,
Phys.  Rev.
Lett. {\bf 83}, 5198 (1999); K.~E.~Strecker, G.~B.~Partridge,
A.~G.~Truscott, and R.~G.~Hulet, Nature {\bf 417}, 150 (2002).

\bibitem{FMS} P. O. Fedichev, A. E. Muryshev,
and G. V. Shlyapnikov, Phys. Rev. A, {\bf 60} 3220 (1999).

\bibitem{Anglin} Th.~Busch and J.~R.~Anglin,
Phys.  Rev.  Lett.  {\bf 84}, 2298 (2000).

\bibitem{BK1} V.~A.~Brazhnyi and V.~V.~Konotop,
Phys. Rev. A {\bf 68}, 043613 (2003).

\bibitem{KP} V. V. Konotop and L. Pitaevskii,
Phys. Rev. Lett. {\bf 93}, 240403 (2004)

\bibitem{PFK} D. E. Pelinovsky, D. J. Frantzeskakis, and P. G. Kevrekidis, Phys. Rev. E {\bf 72}, 016615 (2005) .

\bibitem{com} 
It is to be clarified here that all the papers~\cite{Anglin,BK1,KP,PFK} used different analytical approaches. In particular, the perturbation theories exploited in \cite{BK1}, \cite{KV94} and \cite{PFK} differ with respect to the imposed boundary conditions at $x\to\pm\infty$. They are the conditions of the {\it constant} (i.e. perturbation independent) phase (i.e. $\arg\Psi(x,t)$) \cite{BK1,KV94} and {\it time dependent} (and thus perturbation dependent) phase \cite{PFK}. The difference in complexity of each of approaches as well as their physical relevance can be appreciated to the full extent by studying the convergence of the complete set of the integrals of motion. This last task, however, goes beyond the scope of the present paper. 

\bibitem{KV} V. V. Konotop and L. V\'azquez,
{\em Nonlinear Random Waves.} (World Sceintific, Singapore, 1994).

\bibitem{LL_mech} see e.g. L.~D.~Landau, and E.~M.~Lifshitz, {\it Mechanics} (Pergamon, Oxford, 1960).

\bibitem{footnote} There are also corrections due
to dependence of the two-body scattering amplitude on momenta.
They, however, are small at typical experimental conditions.

\bibitem{Gir} M.~Girardeau, J. Math. Phys. (NY), {\bf 1}, 516 (1960).

\bibitem{Lieb} E. H. Lieb and R. Seiringer, Phys. Rev. Lett. {\bf 91}, 150401 (2003).

\bibitem{Kolom}  E. B. Kolomeisky, T. J. Newman, J. P. Straley,
and Xiaoya Qi, Phys. Rev. Lett. {\bf 85}, 1146 (2000).

\bibitem{Sal}  L. Salasnich, A. Parola and L. Reatto, Phys. Rev. A {\bf 65}, 043614 (2002).


\bibitem{KV94} V.~V.~Konotop and V.~E.~Vekslerchik, Phys.  Rev. E {\bf 49}, 2397 (1994).

\bibitem{Pit}F.~Dalfovo, S.~Giorgini, L.~P.~Pitaevskii, and S.~Stringari,
Rev. Mod.  Phys.  {\bf 71}, 463 (1999).

\bibitem{LL5} L.~D.~Landau, and E.~M.~Lifshitz, {\em Statistical Physics,
Part 1}, (Pergamon Press, Oxford 1986).

\bibitem{BKP} F. G. Bass, V. V. Konotop, and S. A. Pusenko, Phys. Rev. A {\bf 46} 4185 (1992).

\bibitem{BK2} V. A. Brazhnyi and V. V. Konotop, Phys. Rev. E {\bf 72}, 026616 (2005)

\bibitem{K} D. J. Frantzeskakis, N. P. Proukakis, and P. G. Kevrekidis
Phys. Rev. A {\bf 70}, 015601 (2004).

 
\end{thebibliography}
\end{document}